\documentclass[twocolumn,showpacs,preprintnumbers,amsmath,amssymb,prb]{revtex4}

\usepackage{graphicx}
\usepackage{dcolumn}
\usepackage{bm}
\usepackage{color}
\begin{document}

\preprint{}

\title{Room-temperature near-infrared silicon carbide nanocrystalline emitters based on optically aligned spin defects}

\author{A.~Muzha$^{1}$}
\email[These authors contributed equally to this work.]{}
\author{F.~Fuchs$^{2}$}
\email[These authors contributed equally to this work.]{}
\author{N.~V.~Tarakina$^{3,4}$}
\author{D.~Simin$^{2}$}
\author{M.~Trupke$^{5}$}
\author{V.~A.~Soltamov$^{6}$}
\author{E.~N.~Mokhov$^{6}$}
\author{P.~G.~Baranov$^{6}$}
\author{V.~Dyakonov$^{2,3,7}$}
\author{A.~Krueger$^{1,3}$}
\email[E-mail:~]{anke.krueger@uni-wuerzburg.de}
\author{G.~V.~Astakhov$^{2}$}
\email[E-mail:~]{astakhov@physik.uni-wuerzburg.de}

\affiliation{$^1$Institute of Organic Chemistry, Julius-Maximilian University of W\"{u}rzburg, 97074 W\"{u}rzburg, Germany \\
$^2$Experimental Physics VI, Julius-Maximilian University of W\"{u}rzburg, 97074 W\"{u}rzburg, Germany \\
$^3$Wilhelm Conrad R\"{o}ntgen Research Centre for Complex Material Systems (RCCM), Julius-Maximilian University of W\"{u}rzburg, 97074 W\"{u}rzburg, Germany \\ 
$^4$Experimental Physics III, Julius-Maximilian University of W\"{u}rzburg, 97074 W\"{u}rzburg, Germany\\
$^5$Vienna Center for Quantum Science and Technology, Atominstitut, TU Wien, 1020 Wien, Austria\\
$^6$Ioffe Physical-Technical Institute, 194021 St.~Petersburg, Russia\\ 
$^7$Bavarian Center for Applied Energy Research (ZAE Bayern), 97074 W\"{u}rzburg, Germany}

\begin{abstract}
Bulk silicon carbide (SiC) is a very promising material system for bio-applications and quantum sensing. However, its optical activity lies beyond the near infrared spectral window for \textit{in-vivo} imaging and fiber communications due to a large forbidden energy gap. Here, we report the fabrication of SiC nanocrystals and isolation of different nanocrystal fractions ranged from 600~nm down to 60~nm in size.  The structural analysis reveals further fragmentation of the smallest nanocrystals into ca.~10-nm-size clusters of high crystalline quality, separated by amorphization areas. We use neutron irradiation to create silicon vacancies, demonstrating near infrared photoluminescence. Finally, we detect, for the first time, room-temperature spin resonances of these silicon vacancies hosted in SiC nanocrystals. This opens intriguing perspectives to use them not only as \textit{in-vivo} luminescent markers, but also as magnetic field and temperature sensors, allowing for monitoring various physical, chemical and biological processes.      
\end{abstract}

\date{\today}


\maketitle

SiC nanocrystals (NCs) and quantum dots are considered as ideal fluorescent agents for bioimaging applications \cite{Fan:2008cw, Botsoa:2008ct, Somogyi:2014jm, Magyar:2013jb}. They are not toxic as II-VI quantum dots \cite{Derfus:2004dh}, photostable compared to dye molecules \cite{Eggeling:1998kt}, and can be produced on a large scale for a low price \cite{Beke:2012cv, Rao:2012wu}. Apart from that, SiC quantum dots are also suggested for cancer therapy\cite{Mognetti:2010ia}. However, SiC is a wide-bandgap semiconductor and associated photoluminescence (PL) occurs in the ultraviolet or visible spectral ranges \cite{Fan:2012kj}. This is not well suited for \textit{in-vivo} applications because the deepest tissue penetration lies in the near infrared (NIR) spectral window  ($700 - 1400$~nm) \cite{Smith:2009cma}. To overcome this problem, it has been suggested to incorporate appropriate colour centres having optical transitions within the bandgap of SiC \cite{Somogyi:2012gh, Riedel:2012jq, Castelletto:2013jj}. 

The desirable material properties for \textit{in-vivo} bioimaging applications are \cite{Somogyi:2014jm}: (i) non-toxicity and bio-inertness, (ii) from sub-$\mathrm{\mu m}$ size down to molecular size (depending on the application), (iii) NIR absorption and photostable NIR emission, (iv) cost efficient production in large quantities. A natural strategy to fulfil the property (i) is to use group-IV (carbon and silicon) nanomaterials. Indeed, carbon nanotubes (from 1.8~$\mathrm{\mu m}$ down to 50~nm in length) have been applied for deep-tissue imaging \cite{Welsher:2009jc, Hong:2012dc} because of their NIR PL in the spectral range from 900~nm to 1500~nm.  

An alternative approach is based on colour centres in $\mathrm{nanodiamonds}$ with a high PL efficiency \cite{Baranov:2011kz}.  The most prominent example is the nitrogen-vacancy (NV) defect  \cite{Neugart:2007cg, Chang:2008ia, Mohan:2010bf}. Due to their unique optical and spin properties, NV defects can also be used for monitoring various magnetic field- and temperature-dependent processes in biophysical systems \cite{McGuinness:2011ho, Horowitz:2012kp, Kucsko:2013gq}, which is not possible with conventional biomarkers. A disadvantage of the NV defect is that the corresponding PL excitation spectrum lies in the green-yellow part  and hence does not fall into the NIR imaging spectral window. Another promising colour centre in  $\mathrm{nanodiamonds}$---the silicon-vacancy (SiV) defect---can be excited using red laser light  \cite{Neu:2011kj}. It remains thermodynamically stable even in  NCs less than 2~nm in size, as has been found in meteoric samples \cite{Vlasov:2013bu}. However,  the SiV defect has not yet shown spin properties allowing for  sensing applications, as in case of the NV defect. 

Motivated by these exciting results, we have fabricated  SiC NCs containing silicon vacancy ($\mathrm{V_{Si}}$) defects \cite{Comment}. The optimum excitation wavelength of these defects is about 800~nm and they emit in the NIR spectral range \cite{Hain:2014tl}, fulfilling the property (ii) for optimal bioimaging. Remarkably, there is a family of $\mathrm{V_{Si}}$-related defects in SiC with room-temperature optical spin readout  \cite{Falk:2013jq, Kraus:2013di}, allowing for sensing magnetic \cite{Kraus:2013vf} and electric \cite{Falk:2014fh} fields as well as temperature \cite{Kraus:2013vf}. Our top-down approach is suitable for mass production of SiC NCs containing different types of atomic-scale defects.   

\begin{figure}[t]
\includegraphics[width=.45\textwidth]{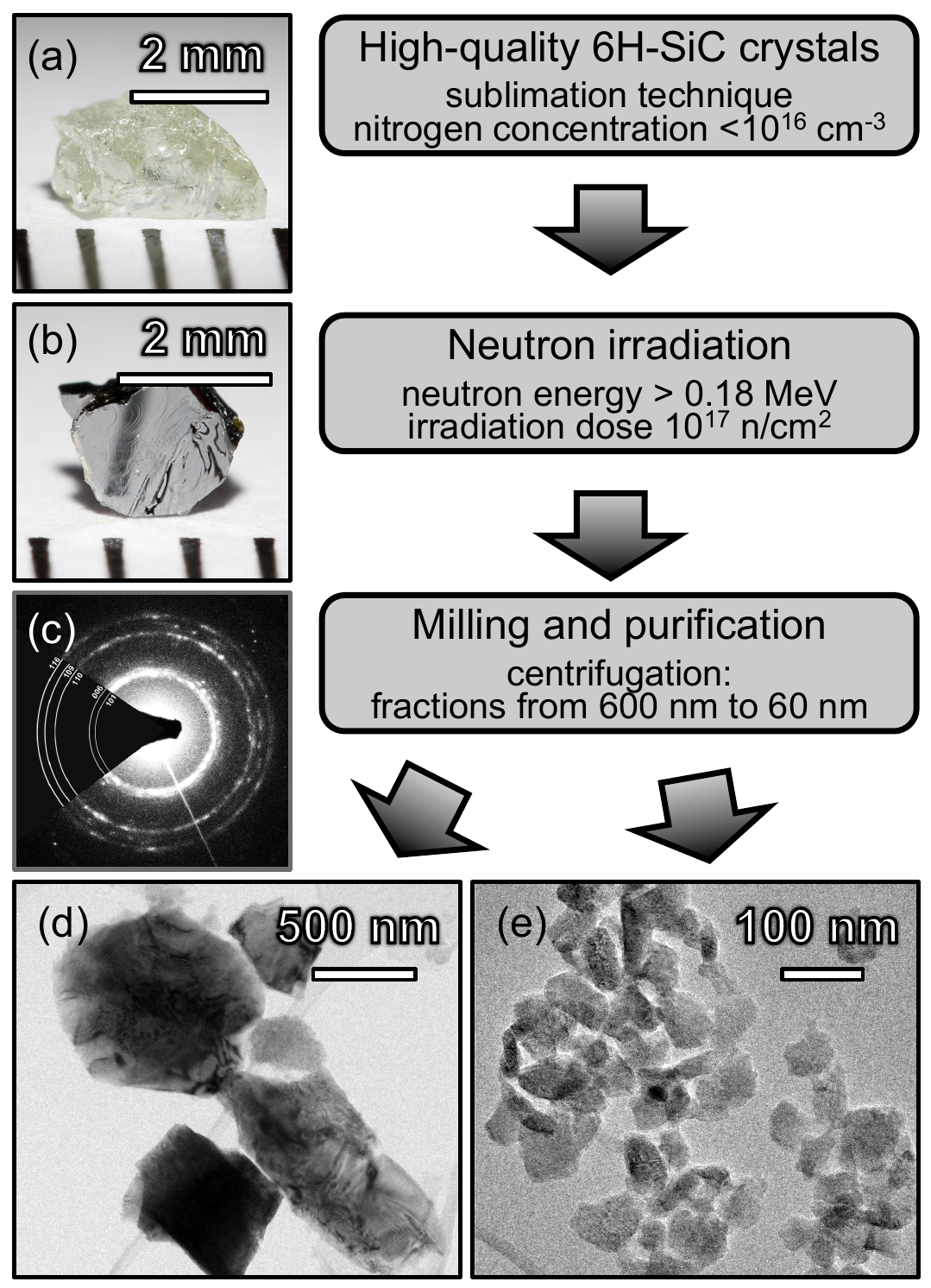}
\caption{Fabrication procedure of SiC NCs. (a) A high-quality 6H-SiC crystal grown by the sublimation technique. (b) A SiC crystal after irradiation with neutrons. The opaqueness is caused by the Rayleigh scattering on irradiation induced defects. (c) Electron diffraction pattern, obtained from the smallest 6H-SiC NC fraction (60~nm). (d), (e) Bright-field TEM images of the SiC NCs of the fractions with average particle size of 600~nm and 60~nm, respectively. } \label{fig1}
\end{figure}

Our fabrication process begins with the growth of high-quality, defect-free SiC bulk material, using the well-established sublimation technique in argon atmosphere at a high temperature (2500 -- 2600~$\mathrm{^{\circ} C}$). The growth was performed from a specially prepared source---pure SiC powder synthesized from  silicon and carbone mixture of spectral purity. Graphite wafers were used as substrates. The high growth temperature allowed to minimize the nitrogen concentration, which is below  $10 ^{16} \, \mathrm{cm ^{-3}}$ in our samples. The final products were mm-size SiC crystals of the polytype 6H [Fig.~\ref{fig1}(a)]. The macroscopic crystal fragments were then placed in the central irradiation tube of a TRIGA Mark-II nuclear reactor, where vacancies were formed by neutron irradiation [Fig.~\ref{fig1}(b)]. The crystals were irradiated for 5 hours 33 minutes to achieve a total dose of fast neutrons of $10^{17} \, \mathrm{cm^{-2}}$. We estimate the concentration of generated $\mathrm{V_{Si}}$ defects to be in the order of $10^{15} \, \mathrm{cm^{-3}}$.

The resulting bulk material was then treated in a high energy milling process that enables the cleavage of lattice bonds and thus allows the production of small particles from large objects. Some of us have recently reported on the production of boron doped $\mathrm{nanodiamonds}$ \cite{Heyer:2014dd}  and $\mathrm{nanodiamonds}$ carrying SiV defects \cite{Neu:2011kj}. The wet-milling process in a vibration mill allows treating small amounts of material and generates a high energy input. Impact along the lattice planes enables the cleaving of brittle materials even down to the nanoscale. Here we used steel milling beads (despite lower hardness compared to SiC).  
The initially very high amount of steel contamination can be reduced by removing most of it using a magnet and finally dissolving the residues in acid. As has been reported before for boron-doped diamond, bond cleavage preferentially occurs at defective lattice sites as bond elongation or lattice distortion result in less stable bonding situations around these defects. It should be therefore considered that the amount of silicon vacancies is reduced in the resulting NCs. From the comparison of the PL intensity shown below, one can roughly estimate that at least 10\% of defects from the initial concentration are still present in the smallest NCs produced in this mechanochemical approach. 

\begin{figure}[t]
\includegraphics[width=.47\textwidth]{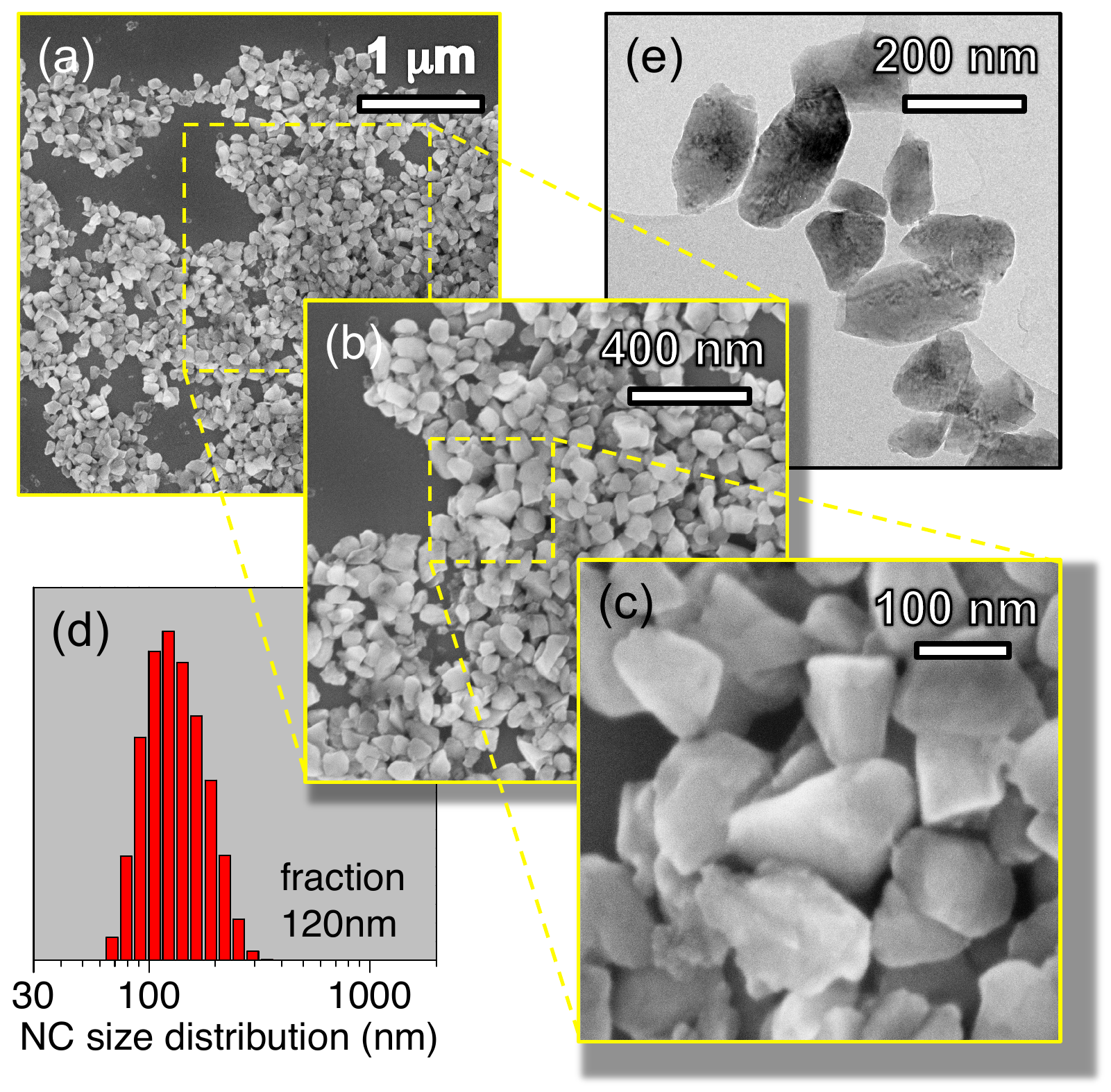}
\caption{Characterisation of the 6H-SiC NC fraction with average particle size of 120~nm (a)-(c) SEM images of the same area obtained at different magnifications. (d) NC size distribution obtained from dynamic light scattering measurements. (e) Bright-field TEM image of a few SiC NCs. } \label{fig2}
\end{figure}

After milling and purification the resulting particle suspension was stabilized using aqueous ammonia to adjust the pH into the stability window of the colloid.  It has to be mentioned that the entire sample is crushed into crystalline particles [Fig.~\ref{fig1}(c)] with diameters well below 1~$\mathrm{\mu m}$. Using fractionated centrifugation, five fractions were obtained with mean NC sizes between 600~nm [Fig.~\ref{fig1}(d)]Ê and 60~nm [Fig.~\ref{fig1}(e)] as aqueous colloids (see also Supporting Information). 

\begin{figure*}[t]
\includegraphics[width=.91\textwidth]{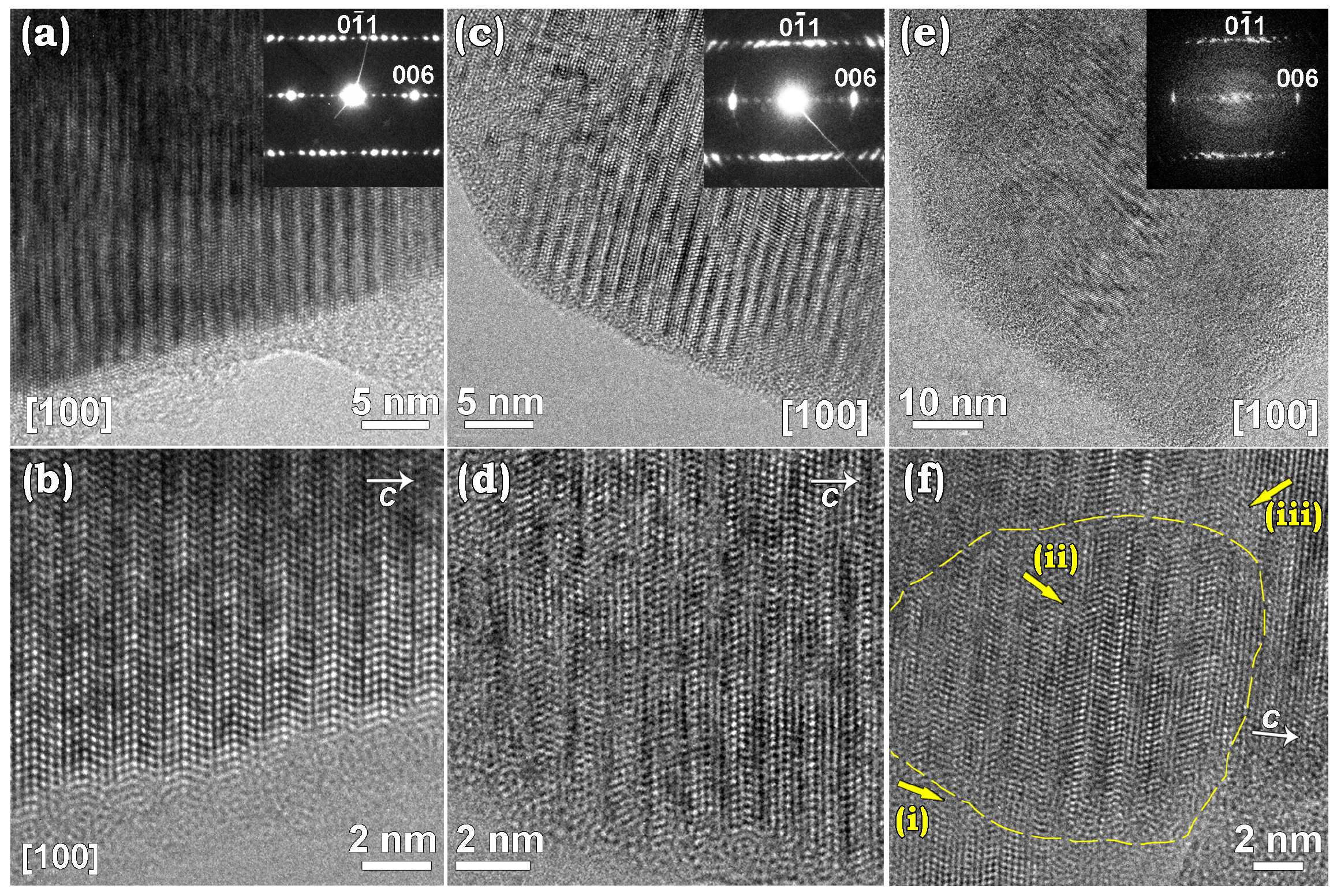}
\caption{HRTEM images and corresponding SAED patterns (inset) of SiC NCs in the fractions with average sizes of 600~nm (a, b), 120~nm (c, d) and 60~nm (e, f), respectively. Note: the inset in (e) shows the fast-Fourier transformation pattern obtained from the crystalline part of the crystal in the image. The dashed line in (f) shows a tentative border of the cluster, drawn along defects and amorphization areas (i) and (iii). The crystal quality within this cluster is comparable with that of bigger crystals; a stacking fault defect is marked by a yellow arrow (ii).} \label{fig3}
\end{figure*}

For scanning electron microscopy (SEM) and PL measurements the samples were prepared by drop casting the aqueous colloid on a silicon wafer and dried at room temperature. Characteristic SEM images of the 120-nm NC fraction are shown in Figs.~\ref{fig2}(a)-(c). The corresponding size distribution was obtained using dynamic light scattering in solution, before casting [Fig.~\ref{fig2}(d)].  In the nominal 120-nm fraction, about 90\% of NCs fall in the range from 80~nm to 250~nm. Remarkably, NCs  exhibit a rather regular but faceted morphology due to the cleavage of bonds along lattice planes. Edges are rounded due to the acid etching during purification. This is clearly seen in a transmission electron microscopy (TEM) image [Fig.~\ref{fig2}(e)], where several NCs are shown. 

Fractions with average NC sizes 600, 120 and 60~nm have been studied in detail using TEM (Fig.~\ref{fig3}). Selected area electron diffraction (SAED) patterns obtained from NCs of different fractions can be indexed in a hexagonal lattice with unit cell parameters $a = 3.08(1)$~{\AA}  and $c = 15.2(1)$~{\AA} [Fig.~\ref{fig1}(c) and Figs.~\ref{fig3}(a), (c) and (e)], indicating that the 6H-SiC structure is preserved during milling and purification. High-resolution TEM (HRTEM) images obtained from the 600~nm fraction show a perfect atomic structure, which corresponds to the 6H-SiC polytype. The only imperfection found in these NCs is a 2~nm amorphous layer at the crystal edges, which is probably caused by acid etching during purification [Fig.~\ref{fig3}(b)]. An amorphous layer of approximately the same thickness is found at the edges of the NCs from the 120~nm fraction [Fig.~\ref{fig3}(d)]. In spite of the fact that the overall quality of the NCs is still very high, the presence of defects is clearly visible from both SAED patterns and HRTEM images. Typical "tailing" of diffraction spots in Fig.~\ref{fig3}(c) suggests that the NCs contain several grains slightly misoriented with respect to each other. The same effect becomes more pronounced in the NCs of the smallest fraction (60~nm). In addition to "tailing" of diffraction spots, which is present in fast-Fourier transformation patterns of Fig.~\ref{fig3}(e), clusters of about 10~nm can be discerned in HRTEM images [Fig.~\ref{fig3}(f)]. The NC quality within such clusters is comparable with that of bigger crystals. We believe that during milling new structural defects tend to appear near existing ones, forming defect areas around highly crystalline areas, which leads to cluster formation. 

\begin{figure}[t]
\includegraphics[width=.49\textwidth]{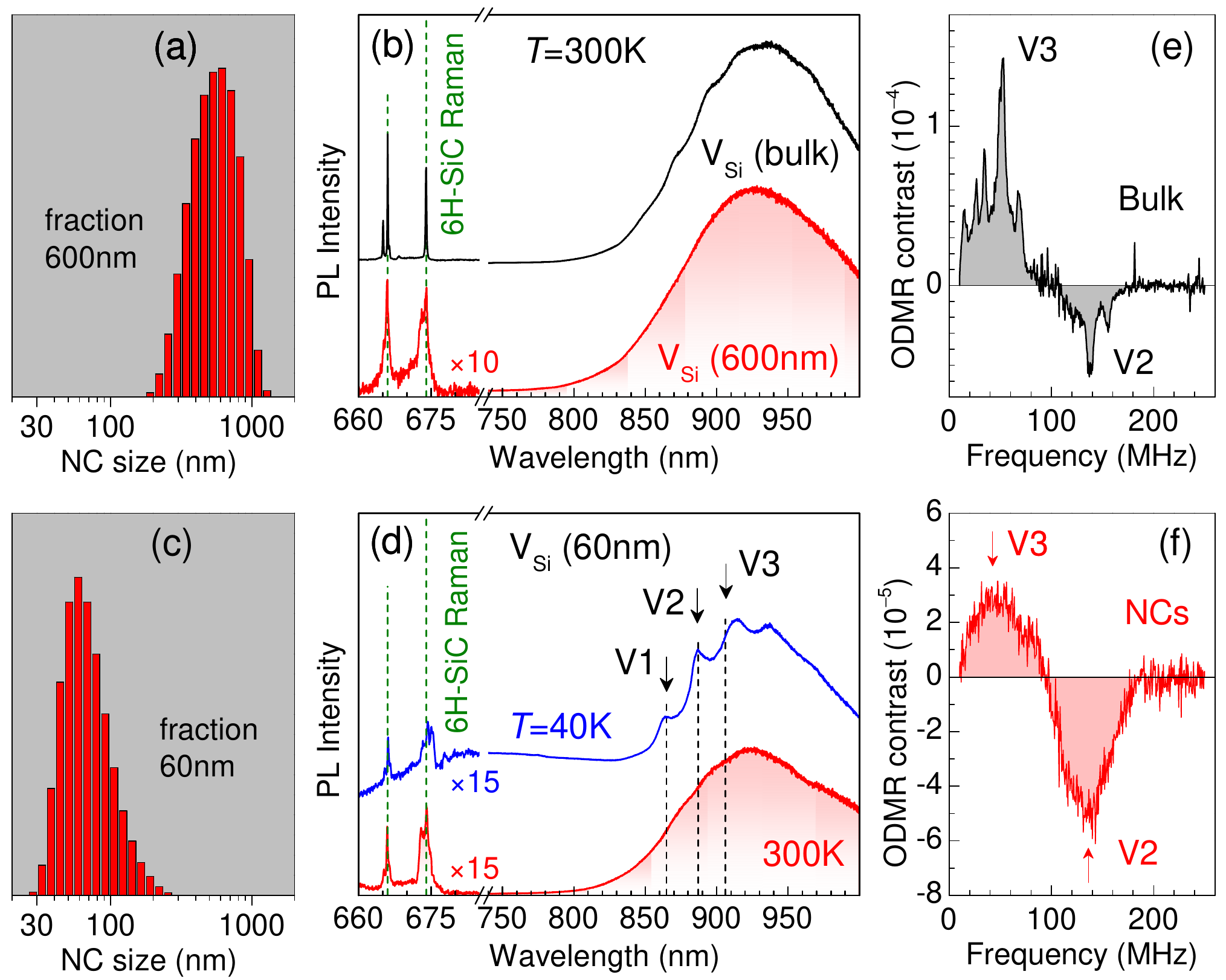}
\caption{NIR PL of SiC NCs  excited at 632.8~nm. (a) NC size distribution in the 600-nm fraction [Fig.~\ref{fig1}(d)]  obtained from the dynamic light scattering measurements. (b) Room-temperature PL spectrum of the $\mathrm{V_{Si}}$ defects within NCs from (a).  A reference PL spectrum of the $\mathrm{V_{Si}}$ defects in bulk 6H-SiC is vertically shifted for clarity. (c) NC size distribution in the 60-nm fraction [Fig.~\ref{fig1}(e)]  obtained from the dynamic light scattering measurements. (d) PL spectra of  the $\mathrm{V_{Si}}$ defects within NCs from (c) recorded at room temperature and at  $T = 40 \, \mathrm{K}$ (vertically shifted for clarity). Arrows indicate the $\mathrm{V_{Si}}$ zero phonon line positions in bulk 6H-SiC:  865~nm (V1), 887~nm (V2) and 906~nm (V3). Spectral position of the 6H-SiC Raman lines in (b) and (d) are shown by dashed lines. (e,f) Room-temperature ODMR spectra of the $\mathrm{V_{Si}}$ defects recorded before milling (e) and in the 600-nm fraction (f).} \label{fig4}
\end{figure}

We now demonstrate that our 6H-SiC NCs are photoluminescent in the NIR at room temperature. We start from the comparison of the largest NCs [600~nm mean size, Fig.~\ref{fig4}(a)] and bulk material [i.e., neutron-irradiated crystals before milling, Fig.~\ref{fig1}(b)]. The samples are excited with a He-Ne laser at  632.8~nm (1.96~eV), far below the  bandgap of 6H-SiC (3.05~eV). The PL spectrum of the bulk crystal consists of a broad emission band from 850~nm to above 1000~nm [Fig.~\ref{fig4}(b)], inherent to the $\mathrm{V_{Si}}$ defects \cite{Fuchs:2013dz}.  It appears after neutron irradiation only and is not observed in as-grown crystals of Fig.~\ref{fig1}(a). The three Raman lines visible in Fig.~\ref{fig4}(b) at shorter wavelength ($664 - 674$~nm) are the spectroscopic Raman fingerprints of polytype 6H. The observed narrow line width is typical for high-quality bulk SiC. We use a confocal microscopy arrangement to find NCs by monitoring the 6H-SiC Raman lines and detect NIR emission in the same spectral range as for bulk SiC [Fig.~\ref{fig4}(b)]. 

We perform this procedure for different fractions and find NIR emission in all of them (see Supporting Information). A typical PL spectrum in the smallest NC fraction [60~nm mean size, Fig.~\ref{fig4}(c)]  is presented in Fig.~\ref{fig4}(d), together with the 6H-SiC Raman lines  to ensure that we probe SiC NCs. To verify that this NIR emission is caused by the $\mathrm{V_{Si}}$ defects, we measure PL at low temperature, where the characteristic zero-phonon lines are clearly silhouetted against the broad phonon sideband emission \cite{Fuchs:2013dz}. Indeed, at $T = 40$~K we observe three peaks, corresponding to the well-established V1 (865~nm), V2 (887~nm) and V3 (906~nm) zero phonon lines of $\mathrm{V_{Si}}$ in 6H-SiC [dashed lines in Fig.~\ref{fig4}(d)] \cite{Wagner:2000fj}. 

Finally, we demonstrate room-temperature optically detected magnetic resonance (ODMR), which constitutes the basis for magnetic field and temperature sensing \cite{Kraus:2013vf}. We measure relative PL difference in a reference sample (before milling) as a function of applied radio frequency and we observe several sharp ODMR lines owing to the zero-field spin splittings in irradiation-induced defects [Fig.~\ref{fig4}(e)].  Remarkably, the V3 and V2 silicon vacancies have spin resonances at $27 \, \mathrm{MHz}$ and  $127 \, \mathrm{MHz}$ with positive and negative ODMR contrast, correspondingly \cite{Kraus:2013vf}. In SiC NCs (the 600~nm fraction), these ODMR resonances are still detectable, while they are significantly broadened [Fig.~\ref{fig4}(f)], presumably because of the variation of zero-field splitting in NCs. 

In summary, we have fabricated near-infrared photoluminescent SiC NCs, using a procedure  compatible with large-scale production. Our starting point was high-quality SiC crystals, grown by the sublimation technique and irradiated with neutrons to generate silicon vacancies, responsible for the NIR emission. 
These crystals have then been milled and dispersed into different NC fractions. The performed HRTEM analysis reveals that the smallest nanocrystals consist of typically 10~nm size clusters of good crystalline quality. 
This indicates that the silicon vacancies  should remain thermodynamically stable even in very small SiC NCs, in agreement with the theoretical prediction \cite{Somogyi:2014jm}.  In additional, we have demonstrated, for the first time, ODMR on vacancy-related defects in SiC NCs at ambient conditions. Given biocompatibility of SiC and that there is a family of hundreds of vacancy-related defects in SiC with distinct quantum properties \cite{Kraus:2013di, Falk:2013jq} allowing for environment sensing \cite{Kraus:2013vf, Falk:2014fh}, our results open attractive opportunities for various quantum sensing applications.  

This work has been supported by the DFG (AS 310/4 and Forschergruppe 1493) as well as by the RFBR (No. 13-02-00821 and No. 14-02-91344). N.V.T. acknowledges funding by the Bavarian Ministry of Sciences, Research and the
Arts.


\begin{thebibliography}{33}
\bibitem{Comment}
\bibinfo{comment}{We note that in case of diamond the silicon-vacancy is an extrinsic defect, consisting of a silicon atom and a vacancy near by. In case of SiC, the silicon vacancy is an intrinsic defect, which is the absence of a silicon atom in the lattice site. In order to avoid confusion we use different notations, i.e., SiV in $\mathrm{nanodiamonds}$ and $\mathrm{V_{Si}}$ in SiC NCs.}
\end{thebibliography}


\end{document}